# Multi-dimension unified Swin Transformer for 3D Lesion Segmentation in Multiple Anatomical Locations

*Shaoyan Pan[1]\*, Yiqiao Liu[2], Sarah Halek[2], Michal Tomaszewski[2], Shubing Wang[2], Richard Baumgartner[2], Jianda Yuan[2], Gregory Goldmacher[2], Antong Chen[2]*

[1] Department of Biomedical Informatics, Emory University, Atlanta, GA, USA
[2] Merck & Co., Inc., Rahway, NJ, USA

*Abstract*—In oncology research, accurate 3D segmentation of lesions from CT scans is essential for the extraction of 3D radiomics features in lesions and the modeling of lesion growth kinetics. However, following the RECIST criteria, radiologists routinely only delineate each lesion on the axial slice showing the largest transverse area, and occasionally delineate a small number of lesions in 3D for research purposes. As a result, to train models to segment the lesions automatically, we normally have plenty of unlabeled 3D volumes, an adequate number of labeled 2D images, and scarce labeled 3D volumes, which makes training a 3D segmentation model a challenging task. In this work, we propose a novel U-shaped deep learning model, denoted a multi-dimension unified Swin transformer (MDU-ST), to automatically conduct 3D lesion segmentation. The MDU-ST consists of a Shifted-window transformer (Swin-transformer) encoder and a convolutional neural network (CNN) decoder, allowing it to adapt to 2D and 3D inputs and learn the corresponding semantic information from various inputs in the same encoder. Based on this model, we introduce a three-stage framework to train the model effectively: 1) leveraging large amount of unlabeled 3D lesion volumes through multiple self-supervised pretext tasks to learn the underlying pattern of lesion anatomy in the Swin-transformer encoder; 2) fine-tune the Swin-transformer encoder to perform 2D lesion segmentation with 2D RECIST slices to learn slice-level segmentation information; 3) further fine-tune the Swin-transformer encoder to perform 3D lesion segmentation with labeled 3D volumes to learn volume-level segmentation information. We compare the proposed MDU-ST with state-of-the-art CNN-based and transformer-based segmentation models using an internal lesion dataset with 593 lesions extracted from multiple anatomical locations and delineated in 3D. The network's performance is evaluated by the Dice similarity coefficient (DSC) for volume-based accuracy and Hausdorff distance (HD) for surface-based accuracy. The average DSC achieved by the MDU-ST with proposed pipeline is 0.78; HD is 5.55 mm. The proposed MDU-ST trained with the 3-stage framework demonstrates significant improvement over the competing models. The proposed method can be used to conduct automated 3D lesion segmentation to assist large-scale radiomics and tumor growth modeling studies.

*Keywords—Lesion segmentation, pre-training, Swin transformer*

## I. INTRODUCTION

Medical imaging applications such as radiomics analysis [1] have the potential to predict lesion-level and patient-level response to oncology treatments based on volumetric lesion-level features [2]. Accurate 3D segmentations of the lesions are required for such analyses. However, radiologists routinely delineate each lesion on one axial slice showing the largest transverse area following the RECIST criteria [3]. Manually delineate the lesions in 3D would require substantial efforts, which is prohibitive when large number of lesions need to be delineated on large-scale studies. An automated segmentation algorithm is thus needed to segment lesions accurately and efficiently in 3D.

Automated image segmentation is currently dominated by architectures based on fully convolutional neural networks (CNN). Among these models, ones based on symmetric U-shaped 2D/3D CNNs, known as the U-Net, have demonstrated effectiveness in various automated lesion/organ segmentation tasks [4]. Despite their success, CNNs have limited abilities to capture long-range spatial relationship, affecting their accuracy in segmentation tasks involving more spatial context. Also, it is impossible for CNN to learn 2D and 3D information simultaneously as it can only use either 2D or 3D inputs for training a 2D or 3D model, respectively.

Vision Transformer [5] (ViT) resolves the above two limitations of CNN models. On the one hand, it has been proposed to capture long-range information to effectively provide more accurate segmentation, especially for objects with varying size and shape in space. On the other hand, we discover that ViT can utilize both 2D and 3D inputs to train the segmentation model. However, ViT is data hungry and often suffers from severe overfitting in medical image segmentation tasks. Among the efforts to reduce overfitting and fully utilize the power of the ViTs, Chen *et al.*[6], Cao *et al.*[7], Pan *et al.*[8, 9] developed a pre-trained transformer, a Swin transformer [10], and token-based transformer layers, respectively, for better multi-organ segmentation accuracy than regular CNNs in 2D/3D medical data. Hatamizadeh *et al.* proposed Swin UNETR [11] which consists of a Swin transformer encoder and a CNN decoder for 3D brain lesion segmentation in MRI.

Herein, we propose a multi-dimension unified Swin transformer (MDU-ST) for lesion segmentation in 3D CT scans using both 2D and 3D images as input. The Swin transformer encoder in the network transforms the input image, regardless of its original dimensions, to a unified 2D embedded feature map with one dimension matching the number of tokens in the input. The learned feature map after transformer encoding can be decoded to the input shape. This property allows us to propose a novel lesion segmentation framework by learning features from the unlabeled 3D dataset and leveraging both 2D and 3D labeled datasets under the same model architecture. Specifically, the proposed segmentation framework consists of three stages. Firstly, we pretrain the encoder by performing contrastive learning and reconstructing masked-out regions from an unlabeled 3D lesion dataset. Secondly, the pretrained encoder is fine-tuned with a set of labeled 2D RECIST slices. Thirdly, the 2D fine-tuned encoder is further fine-tuned with a labeled 3D dataset. The corresponding 2D and 3D decoders consist of 2D and 3D convolutional kernels, respectively. We conducted experiments on an internally collected CT dataset with 593 3D labeled lesions, 23,785 2D labeled RECIST slices and 12,856 unlabeled 3D volumes. Quantitative and qualitative evaluations are included.

## II. METHOD

The MDU-ST network employs a 3D U-shaped encoder-decoder architecture (Fig. 1a) with three stages of training: in Stage 1, based on an unlabeled 3D lesion dataset, we pretrain the encoder by performing contrastive learning with a multi-layer perceptron and reconstructing masked-out regions with a first 3D CNN decoder; in Stage 2, the pretrained encoder is fine-tuned with a 2D CNN decoder using a set of manually labeled 2D RECIST slices; in Stage 3, the 2D fine-tuned encoder from Stage 2 is further tuned with a 3D CNN decoder using a manually labeled 3D dataset. In the proposed three stages, the MDU-ST has the same encoder to store the information from different stages while utilizing different independent decoders for different training tasks. Eventually, the fine-tuned encoder and 3D decoder from Stage 3 is used to perform the 3D tumor segmentation task.

### II.A Encoder architecture

The encoder is a contraction network consisting of one patch embedding layer followed by four Swin transformer down-sample blocks to capture the multi-scale semantic features in the input.

### II.A.1 Patch embedding layer

The patch embedding layer is a down-sampling residual convolutional block consisting of two parallel paths: in the first path, the input is passed into a 3D convolutional layer, which contains 32 convolution filters with kernel size of $3 \times 3 \times 1$ and stride of $2 \times 2 \times 1$, followed by a 3D convolutional layer with an isotropic kernel size of $3 \times 3 \times 3$ and stride of $1 \times 1 \times 1$. The second path has a single 3D convolutional layer with kernel size of $1 \times 1 \times 1$ and stride of $2 \times 2 \times 1$. A shortcut connection [12] is used between the first and the second path to obtain the final down-sampled output.

### II.A.2 Swin transformer down-sample block

Each Swin transformer down-sample block consists of a down-sample convolutional layer, a window self-attention (W-SA) module, and a shifted-window self-attention (SW-SA) module. As shown in Fig. 1 (b), W-SA consists of a window-partitioning layer to divide the input features into non-overlapping windows. Then a multi-head self-attention (MHSA) layer is applied to calculate the global-level spatial information for each window. A linear layer is then applied to embed all the windows into a feature map. The SW-SA has an identical structure to the W-SA module, while a predetermined distance shifts the non-overlapping windows.

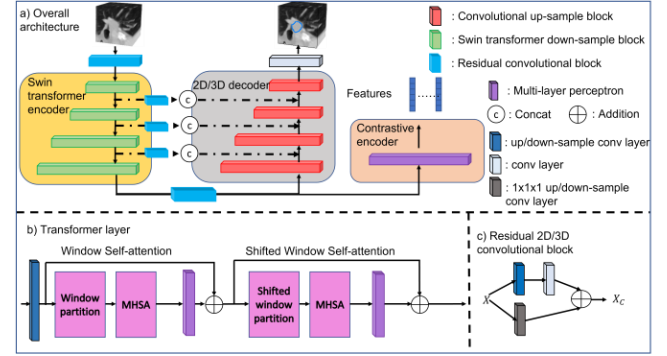

Figure 1. a) The Swin transformer encoder is used for all three stages. Three 2D/3D convolutional decoders with the same structure are used for the volume reconstruction in Stage 1 and the 2D and 3D segmentation in stages 2 and 3. The contrastive encoder is used for contrastive learning in Stage 1. b) The Swin transformer layer consists of a down-sample convolutional layer, a window self-attention and shifted window self-attention. c) The residual convolutional block consists of two pathways to be added.

With the input features $X \in \mathbb{R}^{H \times W \times L \times C}$, we down-sample it, by the down-sampling convolutional layer, to $X_e \in \mathbb{R}^{\frac{H}{2} \times \frac{W}{2} \times L \times 2C}$. In the W-SA module, following the practice in [7], we first partition the embedded feature map into non-overlapping windows by the window partition layer. More specifically, $X_e$ is divided into $\frac{H}{2N} \times \frac{W}{2N} \times \frac{L}{N_L} \times 2C$ windows with the size of $N \times N \times N_L$, where $N$ was empirically set to 4, 4, 8, 4 for the first to the fourth Swin transformer blocks, $N_L$ was set to $N$ for 3D data and set to 1 for 2D data. We denote the partitioned input as a 3D feature map $X_{par} \in \mathbb{R}^{\left(\frac{H}{2N} \times \frac{W}{2N} \times \frac{L}{N_L} \times 2C\right) \times (N \times N \times N_L)}$. For the $l$'th window $M_l$, we utilize a MHSA module comprises of $I$ parallel self-attention heads, each of which learns global features across the window. Each head $i$ consists of independent weight matrices Query (Q), Key (K), and Value (V), which performs:

$$head_i = \text{softmax}(\frac{M_l Q_i (M_l K_i)^T}{\sqrt{d_k}})(M_l V_i)$$

$$A_l = \text{Concat}(head_i, \ldots, head_I) W_{out}$$

where $A_l \in \mathbb{R}^{(N \times N \times N_L)}$ is the concatenation of attentions calculated for the $l$'th window, $W_{out}$ is another weight matrix. By gathering all the $A_l$'s, we finally obtain an attention map $X_{out}$ with the same size as the input $X_e$. An additional linear layer is connected to it for better embedding performance. The final output $X_e^{out}$ will be reshaped back to the original size of the non-partitioned features as $X_e^{out} \in \mathbb{R}^{\frac{H}{2} \times \frac{W}{2} \times L \times 2C}$.

The $X_e^{out}$ is then passed into the SW-SA module, which shifts the partitioned windows by $(\frac{N}{2}, \frac{N}{2}, \frac{N_L}{2})$ voxels for 3D data, and $(\frac{N}{2}, \frac{N}{2}, 1)$ voxels for 2D data. Therefore, the following attention calculation can obtain the attentions across the non-shifted windows in the previous W-SA module. Formally, the Swin transformer layer performs:

$$X_e^{out} = linear(\text{W-SA}(X_e))$$
$$Y_{out} = linear(\text{SW-SA}(X_e^{out}))$$

Where $Y_{out}$ is the final output of the transformer layer.

II.B Decoder architecture

In Stage 1, a reconstruction decoder and a contrastive encoder are connected in parallel to the encoder. The reconstruction decoder is a symmetric expansion network with four 3D up-sampling residual convolutional blocks, as shown in Fig. 1(c), followed by a final convolutional layer with an isotropic kernel size of 1 and stride size of 1. The outputs from the two paths are added to obtain the output feature map. On the other hand, the contrastive encoder consists of two sequential linear layers followed by an instance normalization and a Leaky-ReLu function with output size of 128 to map the encoder's features to a latent representation.

In Stage 2, a 2D segmentation decoder is connected to the encoder. The 2D segmentation decoder consists of four 2D up-sampling residual convolutional blocks to generate a 2D segmentation map from the features. In Stage 3, we connect a 3D segmentation decoder with the same structure as that in Stage 2, just replacing the 2D blocks with 3D blocks and generating a 3D segmentation probability map in the end.

In this network, an instance normalization followed by leaky-ReLu activation is applied after every convolutional layer (except for the final convolutional layer). Furthermore, layer normalizations are applied after every MHSA linear layer except for the last linear layer in the contrastive encoder. To train the networks, we employed the Adam optimizer with a learning rate of 0.0001.

II.B.1 Stage 1: Self-supervised volume reconstruction and contrastive learning

In Stage 1, we pretrain the MDU-ST's encoder with two self-supervised learning tasks, contrastive learning and volume reconstruction, to learn the general 3D information. In the volume reconstruction task, we randomly mask out a Region-of-Interest (ROI) in the lesion volume $X$ with volume ratio of 15 %. The Swin transformer encoder and the reconstruction decoder are trained to recover the masked-out lesion volume to the original volume $X$ with the mean absolute error (MAE) loss function.

Given a lesion volume and its corresponding masked-out volume (positive pair), the contrastive learning trains the network to output representations that maximize the mutual information for the two volumes. For a pair of two different lesions (negative pair), the network is trained to output representations with minimized mutual information by normalized temperature-scaled cross entropy loss (NT-Xent) [13]. Accordingly, the final objective function is a sum of the MAE loss and NT-Xent loss. After the pre-training is finished, the decoders are removed. The encoder carrying general 3D representation is used for the task in the next stage.

II.B.2 Stage 2: Supervised 2D segmentation

In Stage 2, we connect a 2D decoder to the pre-trained encoder to learn 2D segmentation information. We expand an axial dimension to the 2D lesion slices so the input can be regarded as a 3D volume with axial length of 1. Accordingly, the output features from each Swin transformer layer have size of $\frac{H}{2} \times \frac{W}{2} \times 1 \times 2C$, which is then reshaped to $\frac{H}{2} \times \frac{W}{2} \times 2C$. To better convey the high-resolution information to the decoder, the features of each Swin transformer layer are passed into 2D residual convolutional blocks, then connected to the corresponding residual convolutional block on the decoder side via skip connections [14]. The network is optimized by Dice and cross entropy (Dice-CE) loss. After the 2D segmentation is finished, the encoder, with both general 3D representation and supervised 2D representation, is used for the task in the next stage.

II.B.3 Stage 3: Supervised 3D segmentation

In Stage 3, we connect a 3D decoder to the encoder. The output features from each encoder's layer are input to 3D residual convolutional blocks, and then concatenated to the corresponding 3D residual convolutional block via skip connections. The final 3D segmentation network is trained by Dice-CE loss.

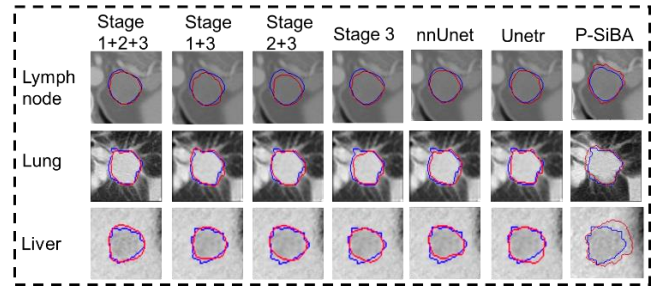

Figure 2. Segmentation results in representative lymph node, lung, and liver lesions from MDU-ST (Stage 1+2+3) and all other competing pipelines and networks. Ground truth and predictions are shown as blue and red, respectively.

II.C Hardware and computational cost

Our model was trained using a single NVIDIA V100 GPU with 32 GB of memory. The batch sizes for training were 16, 16, and 2 for stages 1, 2, and 3, respectively. The model comprises 6,861,859, 8,735,852, and 19,156,124 parameters for stages 1, 2, and 3, respectively. The computational complexity for each stage is as follows: 7.16 GFLOPs for stage 1, 528 MFLOPs for stage 2, and 42.14 GFLOPs for stage 3.

III. DATA PREPROCESSING

To train the network in Stage 1, we utilized an internal CT dataset with 12,865 3D lesion volumes with resampled resolution of 0.75 × 0.75 × 0.75 mm$^3$ with RECIST manual delineations. Extracting the longest diameter $d$ of the lesion from RECIST delineation, we cropped the 3D lesion volume by 2$d$ × 2$d$ × 2$d$ in $x$, $y$, $z$ dimensions with the RECIST delineation at the center. For input to the MDU-ST, the cropped 3D lesion volumes were zero-padded/cropped in the boundary to a size of 128 × 128 × 64. The voxel intensity of the resampled volumes was shifted to positive values by adding 1024 HU and divided by 3000 HU for normalization. To train the network in Stage 2, we aimed to segment the lesions on 23,785 2D RECIST-slices with resampled pixel size of 0.75 × 0.75 mm$^2$. The input 2D RECIST slices were cropped to 2$d$ × 2$d$, padded/cropped to a size of 128 × 128, HU zero-shifted and normalized. For Stage 3, 3D delineations were obtained on 593 internally collected lesions. The same preprocessing was applied on Stage 3 3D volume input as Stage 1. We split the 593 3D lesions into training/validation/testing at a ratio of 416/60/117. For better computational efficiency, the inputs in all stages were down-sampled by a ratio of 2. The network's output probability maps were resampled to the original sizes using trilinear interpolation followed by Softmax and argmax functions to generate the final binary mask. All datasets over the 3 stages contain lymph node, lung, and liver lesions.

IV. PERFORMANCE EVALUATION

We evaluated the proposed network pipeline using the Dice similarity coefficient (DSC) and surface-based Hausdorff distance (HD). To validate the effectiveness of stages 1 and 2 in the pipeline, we conducted ablation study and compared the proposed MDU-ST trained with the full pipeline Stage 1+2+3 with Stage 1+3, Stage 2+3, and Stage 3. In addition, MDU-ST (Stage 1+2+3) was compared with other three state-of-the-art segmentation networks: nnUnet [15], Unet transformer (UNETR) [16], and progressive SIBA (P-SIBA) [17]. The paired t-test with 0.95 confidence interval is applied for statistical significance test.

V. RESULT

Visual comparison of the proposed method with all competing methods is shown in. Fig. 2. MDU-ST (Stage 1+2+3) achieved the best visual results. The quantitative ablation study and comparison with other competing networks in Sec. IV are shown in Table 1.

In terms of the average DSC of lesion segmentations, the full pipeline Stage 1+2+3 demonstrates improvement of about 0.04, 0.03, and 0.05 compared to Stage 1+3, Stage 2+3, and Stage 3. In addition, MDU-ST (Stage 1+2+3) also obtained better DSC over the competing networks: 0.03 over nnUnet, 0.05 over UNETR, and 0.13 over P-SiBA. The proposed MDU-ST (Stage 1+2+3) also demonstrated the lowest standard deviation in DSC, which indicates that the proposed network can achieve more consistent performance across different lesions. In addition, compared to all the competing pipelines and networks, the proposed MDU-ST (Stage 1+2+3) not only obtains better quantitative accuracy, but also obtains p-values < 0.05 in the paired t-test, which demonstrates statistically significant improvement compared to the competing methods.

Table 1: Quantitative and statistical evaluations for the proposed MDU-ST and the competing networks. MDU-ST (Stage 1+2+3) indicates the network with full pipeline; MDU-ST (Stage 1+3) indicates the network with only Stage 1 3D self-supervised pre-training; MDU-ST (Stage 2+3) indicates the network with only Stage 2 2D RECIST supervised training. Paired t-test results are shown comparing the MDU-ST (Stage 1+2+3) with the other networks. The best results are bolded, and the second-best results are underlined.

| Pipelines/ Networks | DSC | Significance | HD (mm) | Significance |
|---|---|---|---|---|
| MDU-ST (Stage 1+2+3) | **0.78±0.08** | N/A | **5.55±2.55** | N/A |
| MDU-ST (Stage 1+3) | 0.74±0.09 | <0.05* | 6.14±2.66 | <0.05* |
| MDU-ST (Stage 2+3) | 0.75±0.09 | <0.05* | 6.33±2.53 | <0.05* |
| MDU-ST (Stage 3) | 0.73±0.11 | <0.05* | 6.73±3.44 | <0.05* |
| nnUnet | <u>0.75±0.09</u> | <0.05* | <u>6.04±2.45</u> | <0.05* |
| UNETR | 0.73±0.10 | <0.05* | 6.69±2.97 | <0.05* |
| P-SiBA | 0.65±0.15 | <0.05* | 17.09±7.51 | <0.05* |

In addition, the MDU-ST (Stage 1+2+3) reduces the average HD by about 0.59 mm, 0.79 mm, and 1.18 mm compared to Stage 1+2, Stage 2+3, and Stage 3. Again, the proposed network also achieved the lowest HD among all the competing networks. Additionally, it obtains one of the lowest standard deviations (only larger than MDU-ST (Stage 2+3) and nnUnet) to demonstrate its consistent performance. The proposed pipeline demonstrates statistically lower surface-based accuracy (all p-values < 0.05) compared to the competing pipelines and networks.

VI. CONCLUSION

This work presents a novel MDU-ST network combining Swin transformer and CNN for automated lesion segmentation in CT images using 2D and 3D inputs. Specifically, we proposed a three-stages pipeline: In Stage 1, we proposed unsupervised contrastive learning and volume reconstruction. Then a supervised 2D segmentation is deployed in Stage 2. Finally, the information gathered from stages 1 and 2 is utilized for supervised 3D lesion segmentation in Stage 3. The approach to train the MDU-ST with Stage 1+2+3 was evaluated on an internally collected set of lesions. It can achieve better performance compared to the MDU-ST trained with only Stage 1+3, Stage 2+3, and Stage 3 alone, which demonstrates the effectiveness of the proposed 3-stage training framework. In addition, it can also achieve better performance compared to other state-of-the-art segmentation networks. The proposed method can be a useful tool for performing automated 3D lesion segmentation to assist large-scale studies in radiomics and tumor growth modeling.